\documentclass[12pt]{article}
\usepackage{amsmath,amssymb}
\usepackage{bm}
\usepackage{color}
\usepackage[dvipsnames]{xcolor}
\usepackage{graphicx}
\usepackage{cite}
\usepackage{mathtools}
\usepackage{breqn}
\usepackage{braket, multirow, subfigure, multicol}
\usepackage{here}
\usepackage{slashed}
\usepackage{comment}
\usepackage{hyperref}

\newcommand{\intZ}{\mathbb{Z}}
\newcommand{\ita}{\mathrm{Im}\,\tau}
\newcommand{\rta}{\mathrm{Re}\,\tau}
\newcommand{\Qbar}{\overline{Q}}
\newcommand{\kbar}{\overline{k}}
\newcommand{\CW}{\mathrm{CW}} 
\newcommand{\Wcal}{\mathcal{W}}

\newcommand{\pr}{\prime}

\newcommand{\GeV}{\mathrm{GeV}}
\newcommand{\QCD}{\mathrm{QCD}}

\newcommand{\ol}[1]{\overline{#1}} 
\newcommand{\order}[1]{\mathcal{O}\left({#1}\right)}

\newcommand{\Ykr}[2]{Y^{({#1})}_{#2}}

\newcommand{\maru}[1]{\raise0.2ex\hbox{\textcircled{\scriptsize{#1}}}}

\numberwithin{equation}{section}

\usepackage{physics}

\begin{document}
\title{
\begin{flushright}
\ \\*[-80pt]
\begin{minipage}{0.23\linewidth}
\normalsize
EPHOU-24-018\\
CTPU-PTC-24-39 \\*[50pt]
\end{minipage}
\end{flushright}
{\Large \bf Large and small hierarchies from finite modular symmetries
\\*[20pt]}}

\author{
~Tetsutaro Higaki$^{a}$,
~Junichiro Kawamura$^{b}$,
~Tatsuo Kobayashi$^{c}$, \\
~Kaito Nasu$^{c}$ and
~Riku Sakuma$^{c}$
\\*[20pt]
\centerline{
\begin{minipage}{\linewidth}
\begin{center}
{\it \normalsize
${}^{a}$Department of Physics, Keio University, Yokohama 223-8533, Japan \\
${}^{b}$Center for Theoretical Physics of the Universe, Institute for Basic Science (IBS), Daejeon
 34051, Korea \\
${}^{c}$Department of Physics, Hokkaido University, Sapporo 060-0810, Japan
}
\\*[5pt]
\end{center}
\end{minipage}}
\\*[50pt]}

\date{
\centerline{\small \bf Abstract}
\begin{minipage}{0.9\linewidth}
\medskip
\medskip
\small
We study the moduli stabilization by the radiative corrections 
due to the moduli dependent vector-like masses invariant under the finite modular symmetry. 
The radiative stabilization mechanism can stabilize the modulus $\tau$ 
of the finite modular symmetry $\Gamma_N$ ($N \in \mathbb{N}$) at $\mathrm{Im}\,\tau \gg 1$, 
where the shift symmetry $\tau \to \tau+1$ remains unbroken approximately. 
The shift symmetry can be considered as the residual $\mathbb{Z}_N$ symmetry 
which realizes the Froggatt-Nielsen mechanism with the hierarchy parameter 
$e^{- 2\pi \mathrm{Im}\,\tau/N} \ll 1$. 
In this work, we study the stabilization of multiple moduli fields, so that 
various hierarchical values of the modular forms coexist in a model. 
For example, one modulus stabilized at $\mathrm{Im}\,\tau_1 \sim 3$ 
is responsible for the hierarchical structure of the quarks and leptons 
in the Standard Model,  
and another modulus stabilized at $\mathrm{Im}\,\tau_2 \sim 15$ 
can account for the flatness of the $\mathrm{Re}\,\tau_2$ direction 
which may be identified as the QCD axion. 
\end{minipage}
}
\begin{titlepage}
\maketitle
\thispagestyle{empty}
\end{titlepage}
\newpage


\section{Introduction}
\label{Intro}

The finite modular symmetries have been intensively studied 
to explain the flavor structure of the quarks and leptons 
in the Standard Model (SM)~\cite{Feruglio:2017spp}.
Under the modular symmetry, coupling constants, typically Yukawa couplings, 
are promoted to the modular forms,  
which are holomorphic functions of modulus $\tau$. 
Some of the finite modular symmetries 
$\Gamma_N$, with level $N \in \mathbb{N}$, are known to be isomorphic 
to the non-Abelian discrete symmetries, such as $\Gamma_3 \simeq A_4$, 
which have been used in bottom-up approaches of flavor model building~\cite{Altarelli:2010gt,Ishimori:2010au,Hernandez:2012ra,King:2013eh,Kobayashi:2022moq}.
It has been shown that the finite modular symmetric models 
can realize masses and mixing of quarks and leptons 
using the modular forms of 
$\Gamma_{2,4} \simeq S_{3,4}$~\cite{Kobayashi:2018vbk,Penedo:2018nmg}, 
$\Gamma_{3,5} \simeq A_{4,5}$~\cite{Feruglio:2017spp,Novichkov:2018nkm}, 
$\Gamma_6$~\cite{Li:2021buv},  
and their double coverings~\cite{Liu:2019khw,Novichkov:2020eep,Liu:2020akv,Wang:2020lxk}~\footnote{
See for reviews Refs.~\cite{Kobayashi:2023zzc,Ding:2023htn}.
}. 
Recently, it has been pointed out that 
the hierarchical flavor structure can be explained 
by the Froggatt-Nielsen (FN) mechanism due to the residual Abelian discrete 
symmetry~\cite{Feruglio:2021dte,Novichkov:2021evw,Petcov:2022fjf,Kikuchi:2023cap,Abe:2023ilq,Kikuchi:2023jap,Abe:2023qmr,Petcov:2023vws,Abe:2023dvr,deMedeirosVarzielas:2023crv,Kikuchi:2023dow,Kikuchi:2023fpl}. 
The FN mechanism is realized when the modulus is stabilized 
nearby one of the fixed points, $\tau = i, e^{2\pi i/3}$ or $i\infty$. 
In particular, the residual $\intZ^T_N$ symmetry, 
corresponding to $T$ transformation $\tau \to \tau+1$, 
remains unbroken if the modulus is stabilized at $\ita \gg 1$.

The stabilization of the modulus is important for the flavor structure. 
In general, the moduli stabilization is a critical problem 
in models with extra dimensions, such as superstring theory, 
where the modulus describes the geometry of extra dimensions.
The moduli stabilization by the modular invariant non-perturbative superpotential 
in 4D low-energy effective field theory of superstring theory 
is discussed in Refs.~\cite{Font:1990nt,Ferrara:1990ei,Cvetic:1991qm}.
Recently, the stabilization scenarios are revisited for modular flavor models 
using the modular invariant function and the modular forms coupled to a matter field 
in Refs.~\cite{Kobayashi:2019xvz,Kobayashi:2019uyt,Novichkov:2022wvg,Knapp-Perez:2023nty,King:2023snq,Leedom:2022zdm,Abe:2024tox,Higaki:2024pql,Fan:2024jvz}.
The stabilization by the three-form flux backgrounds~\cite{Gukov:1999ya} 
is also studied in Ref.~\cite{Ishiguro:2020tmo,Ishiguro:2022pde}.
Interestingly, these modulus potentials may be applicable for the inflation in the early universe~\cite{Kobayashi:2016mzg,Abe:2023ylh,Ding:2024neh,King:2024ssx,Casas:2024jbw}. 
Also, 
it is found in Ref.~\cite{Kikuchi:2023uqo} that 
the moduli dependent mass of matter fields 
will result in the moduli trapping behavior.

The radiative corrections may also be important to stabilize the modulus. 
The Coleman-Weinberg (CW) potential can deviate the modulus from a fixed point 
stabilized by a tree-level potential~\cite{Kobayashi:2023spx}.
It is shown in Refs.~\cite{Higaki:2024jdk,Jung:2024bgi} that 
the CW potential from the mass terms alone can stabilize the modulus at $\ita \gtrsim \order{1}$, 
where the residual $\intZ^T_N$ symmetry is unbroken, 
if the mass is given by a certain modular form whose $\intZ^T_N$ charge is non-zero.  
Interestingly, the potential along $\rta$ direction is very flat 
due to the accidental $U(1)$ symmetry originated from the residual $\intZ^T_N$ symmetry. 
The accidental $U(1)$ symmetry could be identified as the PQ symmetry  
so that the strong CP problem is solved by the QCD axion $\sim \rta$~\cite{Higaki:2024jdk}~\footnote{
The strong CP problem may also be solved by forbidding the CP violating $\theta$ angle
by the finite modular symmetry~\cite{Kobayashi:2020oji,Feruglio:2023uof,Petcov:2024vph,Penedo:2024gtb,Feruglio:2024ytl,Feruglio:2024dnc}.
}. 
The majoron is realized if the accidental $U(1)$ is identified as the global $U(1)_{B-L}$ symmetry~\cite{Jung:2024bgi}.  
Typically, $\ita \gtrsim \order{10}$ is required 
to realize the very flat direction in the potential to be identified as the axion, 
while $\ita \sim \order{3}$ is preferable to explain 
the hierarchical structure of the quarks and leptons. 
For example, $\ita\sim 3$ is favored to explain the flavor hierarchies under $\Gamma_6$~\cite{Kikuchi:2023cap,Abe:2023dvr}. 
Thus, we may need more than one modulus to address the problems 
in the SM in a model.

In this paper, we extend the radiative stabilization scenario 
in Ref.~\cite{Higaki:2024jdk} to the multiple moduli case. 
We consider the direct product 
of the finite modular symmetry 
$\Gamma_{N_1} \times \Gamma_{N_2} \times \cdots \times \Gamma_{N_n}$, 
where each symmetry has its modulus $\tau_i$~\footnote{
It will be interesting to consider the Siegel modular group 
to stabilize multiple modulus fields, 
but this is not a scope of this paper. 
The flavor structures form the Siegel modular group 
are discussed in Refs.~\cite{Kikuchi:2023awe,Ding:2020zxw,Ding:2021iqp,RickyDevi:2024ijc,Ding:2024xhz,Kikuchi:2023dow}, 
and the moduli stabilization is studied in Ref.~\cite{Funakoshi:2024yxg}.
}. 
We first discuss some necessary conditions to stabilize the moduli 
in the model, 
and then we demonstrate the stabilization of two moduli 
under $\Gamma_6 \times \Gamma_6$ as an illustrating example. 
The flatness of $\rta_i$ direction is also discussed.

This paper is organized as follows.
We briefly review the finite modular symmetry in Section ~\ref{sec:modular}. 
In Section~\ref{sec:radiative}, 
we discuss the moduli stabilization by the CW potential. 
Section~\ref{sec:conclusion} is devoted to conclusion.
In Appendix~\ref{appendix:gamma6}, 
we show the modular forms of $\Gamma_6$ used in our analysis.


\section{Finite modular symmetry}
\label{sec:modular}
In this section, we briefly review the modular symmetry.
We consider the modular group ${\rm SL}(2,\mathbb{Z})$ consisting of 
the $2\times 2$ matrices parametrized as  
\begin{align}
\gamma = 
    \begin{pmatrix}
        a & b \\ c & d
    \end{pmatrix},
\end{align}
where $a,b,c,d$ are integers satisfying $ad-bc=1$.
This modular group is generated by three generators
\begin{equation}
S=\begin{pmatrix}
0 & 1 \\
-1 & 0 \\
\end{pmatrix},~
T=\begin{pmatrix}
1 & 1 \\
0 & 1 \\
\end{pmatrix},~
R=\begin{pmatrix}
-1 & 0 \\
0 & -1 \\
\end{pmatrix}.
\end{equation}
The modulus $\tau$ ($\ita > 0$) transforms under the modular transformation $\gamma \in \mathrm{SL}(2,\mathbb{Z})$,
\begin{equation}
\tau\rightarrow \gamma \tau = \frac{a\tau+b}{c\tau+d}.
\end{equation}
We then introduce the congruence subgroup:
\begin{equation}
\Gamma(N):=\left\{\begin{pmatrix}
a & b \\
c & d \\
\end{pmatrix}
\in \mathrm{SL}(2, \mathbb{Z}), 
\quad 
\begin{pmatrix}
a & b \\
c & d \\
\end{pmatrix}
\equiv
\begin{pmatrix}
1 & 0 \\
0 & 1 \\
\end{pmatrix}
~~\mathrm{mod}~N
\right\},
\end{equation}
where the integer $N$ is called level. 
The finite modular symmetry is defined as the quotient group $\Gamma_N:=\bar{\Gamma}/\Gamma(N)$ where $\bar{\Gamma}:=\Gamma(1)/\mathbb{Z}_{2}^{R}$. Here $\mathbb{Z}_{2}^{R}$ is the $\mathbb{Z}_2$ symmetry generated by $R$.
The modular form, 
with the representation $r$ under $\Gamma_N$ and the modular wight $k\in \mathbb{Z}$, 
is the holomorphic function of $\tau$, which transforms as  
\begin{equation}
\Ykr{k}{r}(\tau) \to \Ykr{k}{r}(\gamma \tau) = (c\tau+d)^k\rho(r)Y_{r}^{(k)}(\tau),
\end{equation}
where $\rho(r)$ is the representation matrix of $\Gamma_N$.
We study models with supersymmetry 
so that Yukawa couplings in the superpotential are identified as the modular forms, 
which are holomorphic functions of the modulus $\tau$.  
We assume that the chiral superfield $Q$ 
with weight $-k^i_Q$ and representation $r^i_Q$ transforms as
\begin{equation}
Q \xrightarrow{\gamma} \left(\prod_i (c_i\tau_i+d_i)^{-k^i_Q}\rho(r^i_Q)\right) Q,
\end{equation}
under the modular transformation. 

\section{Radiative moduli stabilization}
\label{sec:radiative}

We study the finite modular symmetric model 
with the pairs of chiral matter fields $(Q_a, \Qbar_a)$ 
which are vector-like under the SM gauge group. 
In this paper, we will consider a direct product of the finite modular groups 
$ \Gamma_{N_1} \times \Gamma_{N_2} \times \cdots \times \Gamma_{N_n}$ 
where each $\Gamma_{N_i}$ has its modulus $\tau_i$ with $i=1,2,\cdots,n$. 
This setup with $n\le 3$ is motivated by the toroidal compactification of the 6D extra-dimension space 
$T^2 \times T^2 \times T^2$,
where each torus may has its $\Gamma_{N_i}$.

We study the simple supergravity model 
with the following K\"ahler potential and superpotential:
\begin{align}
\label{eq-genKW}
K=&\ - \sum_i h_i \log(-i\tau_i+i\tau_i^\dagger)
   +\sum_a
  \left(\frac{    Q_{a}^{\dag}  Q_a   }{\prod_i(-i\tau_i+i\tau_i^\dag)^{k^i_{Q_a}}}
       +\frac{\Qbar_{a}^{\dag} \Qbar_a}{\prod_i(-i\tau_i+i\tau_i^\dag)^{\kbar^i_{Q_a}}}
 \right),
\\
W=&\ \sum_{a} \Lambda_{Q_a} 
 \left[
  \left(\prod_i \Ykr{k^i_a}{r^i_a}(\tau_i)\right) \bar{Q}_aQ_a 
\right]_{1_0},
\end{align}
where $h_i\in\mathbb{N}$ and $\Lambda_{Q_a}$ is the mass scale 
of the vector-like mass for $(Q_a, \Qbar_a)$. 
Here, $r^i_a$ and $k^i_a$
are respectively the representation and the modular weight 
of the modular form 
under the finite modular symmetry $\Gamma_{N_i}$. 
We use the unit of $M_p=1$ with $M_p$ being the reduced Planck scale.
In general, 
we take the trivial-singlet combinations of the product denoted by $[\cdots]_{1_0}$, 
but we focus on the case of the singlets $1_t$, with $t$ being $\intZ^T_N$ charge,  
and no flavor mixing for simplicity~\footnote{
Under $\Gamma_6$ which will be used in our numerical analysis, 
there are singlet representations $1_t$ with $t=0,1,2,\cdots,5$. 
In the notation in Ref.~\cite{Li:2021buv}, they correspond to   
$1_0 = 1^0_0, 1_1 = 1^1_2, 1_2 = 1^0_1, 1_3 = 1^1_0, 1_4 = 1^0_2, 1_5 = 1^1_1$, 
where the upper (lower) index is the charge under the $\intZ_2$ ($\intZ_3$)   
symmetry of $\intZ_6^T \simeq \intZ_2 \times \intZ_3$. 
See Refs.~\cite{Kikuchi:2023cap,Abe:2023dvr,Li:2021buv} for the details of $\Gamma_6$ and its double covering $\Gamma_6^\prime$. 
}. 
In this case, the charges under the $\intZ^T_N$ symmetry 
and the modular weights satisfy 
\begin{align}
 t^i_a + t^{i}_{Q_a} + \ol{t}^{i}_{Q_a} \equiv 0, 
\quad  
 k^i_a = - h_i + k^i_{Q_a} + \ol{k}^i_{Q_a}, 
\end{align}
for any $a$'s,
so that $G:= K + \log \abs{W}^2$ is invariant under the modular symmetry. 
Here, the $\intZ^T_{N_i}$ charges of the modular form, $Q_a$ and $\Qbar_a$ 
are defined as $t^i_a$, $t^i_{Q_a}$, $\ol{t}^i_{Q_a}$, respectively, and 
$\equiv$ is understood as $\mathrm{mod}~N_i$.  
First, we study the single modulus case $n=1$, 
then the multi-moduli case is discussed later.

\subsection{Single modulus case} 

We first discuss the single modulus case, so we omit the index $i$ in this section.   
The vector-like mass terms in the superpotential induce the Coleman-Weinberg (CW) potential:
\begin{align}
\label{eq:CW-V}
V_{\mathrm{CW}}=&\ 
 \frac{1}{32\pi^2}\sum_a\left[(m_{a}^{2}+m_{Q_a}^{2}(\tau))^2
 \left\{\log\left(\frac{m_{a}^{2}+m_{Q_{a}}^{2}(\tau)}{\mu^2}\right)-\frac{3}{2}\right\}
\right. 
\notag \\
& \left. \hspace{2.0cm} 
-(m_{Q_a}^{2}(\tau))^2
\left\{\log\left(\frac{m_{Q_a}^{2}(\tau)}{\mu^2}\right)-\frac{3}{2}\right\}\right], 
\end{align}
where the first and second terms are 
due to bosonic and fermionic contributions, respectively.
Here, we consider the $\overline{\rm MS}$ scheme 
and $\mu$ denotes the renormalization scale.
The canonically normalized mass square of the vector-like fermion $(Q_a, \Qbar_a)$ 
is given by 
\begin{equation}
\label{eq:mass-canonical}
m_{Q_a}^{2}(\tau)=\Lambda_{Q_a}^{2}(-i\tau+i\tau^\dagger)^{k_a}
                  \abs{\Ykr{k_a}{1_{t_a}}(\tau)}^2, 
\end{equation}
and the scalar mass squared is given by $m_a^2 + m_{Q_a}^2(\tau)$, 
where $m_a^2$ is the soft supersymmetry breaking mass squared.   
For simplicity, 
we assume that $m_a^2$ is independent of $\tau$~\footnote{
See Ref.~\cite{Kikuchi:2022pkd}, 
for discussions about supersymmetry breaking parameters 
transformed under the modular symmetry.}.

The potential has a minimum at $\ita \gg 1$ 
as shown in Refs.~\cite{Higaki:2024jdk,Jung:2024bgi}. 
To illustrate this, we start from the simplest case with only one vector-like pair, 
so the index $a$ is omitted hereafter. 
We parametrize the potential as 
\begin{align}
 V_{\mathrm{CW}} =: \frac{\Lambda_{Q}^4}{16\pi^2}\; U(X),
\end{align} 
with
\begin{align}
 U(X) := \frac{1}{2} \left[\left(\alpha + X\right)^2 \left\{
 \log\left( \alpha + X \right) + L_\mu -\frac{3}{2}  
  \right\} 
- X^2 \left( \log X + L_\mu - \frac{3}{2} \right)\right], 
\end{align}
where $\alpha := m^2/ \Lambda_Q^2$, $L_\mu := \log \Lambda_Q^2/\mu^2$ and 
\begin{align}
 X = X(\tau) := (-i\tau+i\tau^\dag)^k \abs{\Ykr{k}{1_t} (\tau)}^2   
\end{align}
is the fermion mass squared normalized by $\Lambda_Q^2$. 
The first derivative of $U(X)$ with respect to $X$ is given by 
\begin{align}
\label{eq-Upr}
 U^\prime(X) =&\ (\alpha+X) \log(\alpha + X) 
                   - X\log X + \alpha (L_\mu - 1) 
= \alpha \log \frac{m_Q^2(\tau)}{\mu^2} + \order{\alpha^2},  
\end{align} 
where $\alpha \ll 1$ is assumed in the last equality. 
Thus the CW potential may have the minimum where 
the vector-like mass $m^2_Q(\tau)$ is at the renormalization scale $\mu^2$.  
The second derivatives of the normalized potential $U$ at this point is given by 
\begin{align}
\begin{pmatrix}
 \partial_{\rta}^2 & \partial_{\rta}  \partial_{\ita} \\
\partial_{\rta}  \partial_{\ita} & \partial_{\ita}^2 \\
\end{pmatrix}
U(X) 
= 
 \log \left(1+ \frac{\alpha}{X} \right)
\begin{pmatrix}
 (\partial_{\rta}X)^2 & (\partial_{\rta} X) (\partial_{\ita}X) \\
 (\partial_{\rta}X)(\partial_{\ita}X) & (\partial_{\ita} X)^2 \\
\end{pmatrix}.  
\end{align}
Since this matrix is rank one, 
it has one massless mode and positive mass along the massive direction.   
In Ref~\cite{Higaki:2024jdk}, 
the massless mode is identified as the QCD axion, 
so that the massless mode 
gets massive due to the QCD effect and the minuscule contribution
from the sub-dominant CW potential. 
We also note that 
the extremum with $dX/d\tau=0$ and $U^\prime(X) \ne 0$  
is the maximum~\cite{Higaki:2024jdk}.

\begin{figure}[t]
\centering
   \includegraphics[width=0.6\hsize]{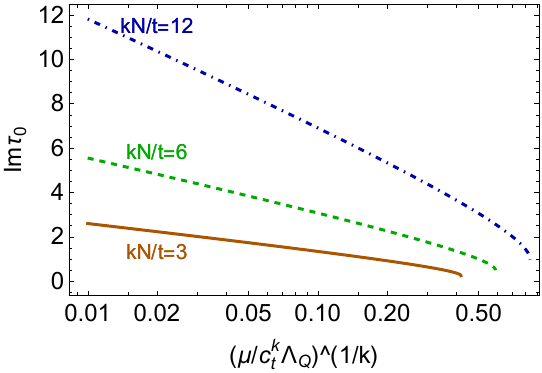}
  \caption{Approximated values of $\ita$ from Eq.~\eqref{eq-ImTauApp}.  }
   \label{fig:ImTauApp}
\end{figure}

The potential has a minimum at large $\ita$ 
for $m^2 \ll \mu^2 \ll \Lambda_Q^2$
as shown in Eq.~\eqref{eq-Upr}.   
For $\ita \gtrsim 1$, the singlet modular form can be expanded as 
\begin{align}
\label{ex-qexp}
 \Ykr{k}{1_t}(\tau) = c^k_t q^{t/N} \left(1 + b_1 q + b_2 q^2 + \cdots  \right),
\end{align}
where $q := e^{2\pi i \tau}$. 
The real coefficients $c^k_t$ and $b_\ell$ ($\ell = 1,2\cdots$) 
are determined by the explicit form of $\Ykr{k}{1_t}$.  
Hence, the normalized mass squared $X$ is given by 
\begin{align}
 X =&\ (c_t^{k})^2 (2\ita)^k e^{-\frac{4\pi t\ita}{N}} 
     \left( 1 + 2b_1 \abs{q} \cos\left(2\pi \rta\right)  
             + \order{\abs{q}^2} \right), 
\end{align}
where $\abs{q} = e^{-2\pi\ita} \ll 1$. 
Hence the solution of $U^\prime(X)=0$ in Eq.~\eqref{eq-Upr} is formally given by 
\begin{align}
\label{eq-ImTauApp}
 \ita_0 := - \frac{kN}{4\pi t} \Wcal \left( 
 -\frac{2\pi t}{kN} \left(\frac{\mu}{c_t^k\Lambda_Q}\right)^{2/k}
  \right) + \order{\abs{q}, \alpha},
\end{align}
where $\Wcal$ is the Lambert W function satisfying $\Wcal(z)e^{\Wcal(z)} = z$
and we take one of the branches with a negative value.
Typical values of $\ita_0$ are shown in Fig.~\ref{fig:ImTauApp}, 
where the value is larger for larger $kN/t$ or smaller $(\mu/c^k_t\Lambda_Q)^{1/k}$. 
Around this minimum, the potential is expanded as 
\begin{align}
U(X) = U(X_0) + U^\prime(X_0) (X(\tau)-X_0) + \order{\abs{q}^2},
\end{align}
where $X_0 := X(\ita_0)$. 
Since $U^\prime(X_0)=0$, 
$\rta$ dependence of the potential appears at $\order{\abs{q}^2}$ 
if there is only one vector-like term is involved to generate the potential.   
While if there are more than one term as discussed in Ref.~\cite{Higaki:2024jdk}, 
$\rta$ dependence appears at $\order{\abs{q}}$ 
from the second dominant vector-like mass. 

  \begin{figure}[t]
  \begin{minipage}{0.49\hsize}
 \centering
   \includegraphics[width=\hsize]{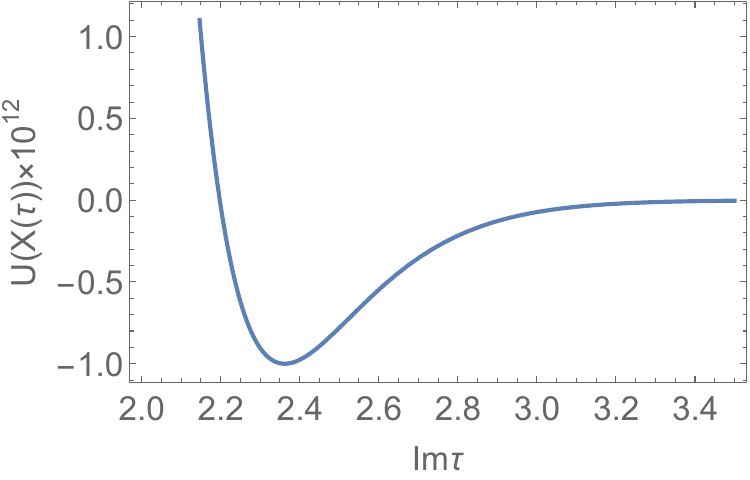}
 \end{minipage}
  \begin{minipage}{0.49\hsize}
 \centering
   \includegraphics[width=\hsize]{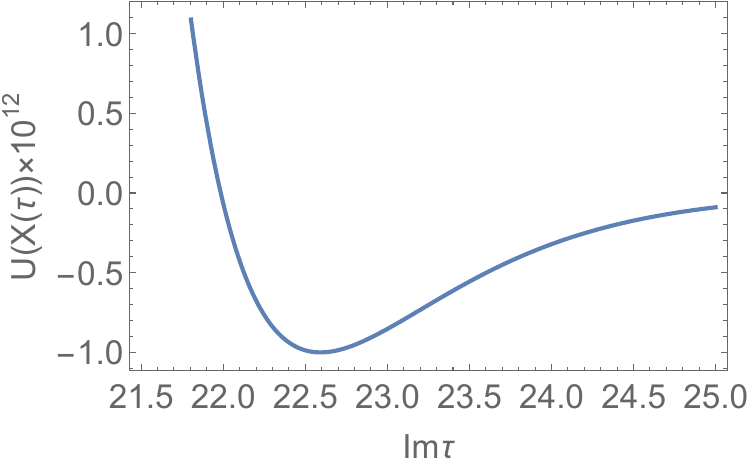}
 \end{minipage}
  \caption{
The shape of the potential along $\ita$ direction at $\rta = 0$ 
for $t=5$ (left) and $t=1$ (right) with $k=10$, 
$\mu/\Lambda_Q = 10^{-2}$ and $m^2/\Lambda_Q^2 = 10^{-8}$. 
The minimums are at $\ita_0\simeq 2.4$ (left) and $\ita_0 \simeq 22.6$ (right). 
}
  \label{fig:weight10}
   \end{figure}
For illustration, we consider the case of $N=6$. 
The explicit forms of the singlet modular forms of $\Gamma_6$
are shown in Appendix \ref{appendix:gamma6}.
Figure~\ref{fig:weight10} shows the potential along the $\ita$ direction at $\rta = 0$, 
but the potential is extremely flat along the $\rta$ direction. 
The left (right) panel is the potential generated 
from the modular forms~\footnote{
Note that $k=10$ is the smallest weight which has the non-trivial singlet $1_5 = 1^1_1$, 
see Appendix~\ref{appendix:gamma6}. 
}, 
$\Ykr{10}{1_5}$ ($\Ykr{10}{1_1}$),  
which realizes its minimum at $\ita_0=2.4$ ($\ita_0 = 22.6$). 
The ratios of the scale parameters are chosen to 
$\mu/\Lambda_Q=10^{-2}$ and $m^2/\Lambda_{Q}^{2}=10^{-8}$.
Thus we see that the value of $\ita$ at the minimum 
can be $\order{1\mathrm{-}10}$ by choosing the representation and weight. 
For the flavor structure, the minimal unit for the hierarchy 
is $\epsilon := \abs{q}^{1/N}$. 
For $N=6$, $\ita \sim 3$ is moderate to explain the hierarchies 
in the SM quarks and leptons~\cite{Kikuchi:2023cap,Abe:2023dvr}. 
While it can be used to realize the axion-like direction 
for $\ita \sim \order{10}$, 
so that it contributes to the dark matter 
and/or
solves the strong CP problem~\cite{Higaki:2024jdk,Jung:2024bgi}.      
These hierarchical structures supported by the finite modular symmetry 
would be applied for generating other scales much smaller 
than a fundamental scale, such as the Planck scale or string scale, 
or perhaps for explaining the tiny cosmological constant.  
To address more than one hierarchy issues in the SM at the same time, 
e.g. explain both flavor hierarchy and strong CP problem in a model, 
we have to realize two different hierarchies at once. 
Another possibility is that the coexistence of the two hierarchies 
may allow to explain the flavor hierarchy and spontaneous CP violation 
due to the Froggatt-Nielsen mechanism~\cite{Kikuchi:2023fpl}.   
We will demonstrate this by simply extending the radiative stabilization scenario 
with multi-moduli.

\subsection{Multi moduli case}
\label{sec:multi-moduli}

Now, we discuss the stabilization mechanism for the multiple moduli, i.e. $n>1$.  
The CW potential generated from the general setup in Eq.~\eqref{eq-genKW} 
only with the singlet representations are the same form as Eq.~\eqref{eq:CW-V},  
but the vector-like masses are generalized to 
\begin{align}
 m_{Q_a}^2(\tau) =  \Lambda_{Q_a}^2 
     \prod_i \left(2\ita_i \right)^{k^i_a} \abs{\Ykr{k^i_a}{1_{t^i_a}}(\tau_i)}^2 
     =: \Lambda_{Q_a}^2 X_a(\tau), 
\end{align} 
where $a=1,2,\cdots,A$ runs over the vector-like pairs. 
Without loss of generality, 
we can take $\Lambda_{Q_a} = \Lambda_Q$ 
by absorbing it to the normalization of the modular forms. 
Then, we define the normalized potential as 
\begin{align}
 V_{\CW}(\tau) =: \frac{\Lambda_Q^4}{16\pi^2} U(\tau), 
\quad 
 U(\tau)  =: \sum_{a=1}^A U_a(X_a),    
\end{align}
with 
\begin{align}
 U_a(X) := \frac{1}{2} \left[\left(\alpha_a + X_a\right)^2 \left\{
 \log\left( \alpha_a + X_a \right) + L_\mu -\frac{3}{2}  
  \right\} 
- X^2_a \left( \log X_a + L_\mu - \frac{3}{2} \right)\right],  
\end{align}
where $\alpha_a := m_a^2/\Lambda_Q^2$. 
The first derivative is given by 
\begin{align}
\label{eq-dtauiU}
 \partial_{\tau_i} U(X) =: \sum_a 
      U_a^\pr \partial_{\tau_i} X_a(\tau) 
  =: \sum_a C_{ia}(\tau_i) \frac{dU_a}{d\log X_a},
  \end{align}
where the derivative $U_a^\pr:= dU_a/dX_a$ can be obtained 
by formally replacing $\alpha \to \alpha_a$, $X \to X_a$ in Eq.~\eqref{eq-Upr}.  
Here, we define the $n\times A$ matrix $[C]_{ia}  $ with 
\begin{align}
 \partial_{\tau_i} X_a(\tau) = \left( \partial_{\tau_i} 
        \log \Ykr{k^i_a}{1_{t^i_a}}(\tau_i) -\frac{ik^i_a}{2\ita_i} \right) 
        X_a(\tau) =: C_{ia}(\tau) X_a(\tau),  
\end{align}
which is approximately given by 
\begin{align}
 C_{ia}(\tau_i)
 = 
 \frac{2\pi i t^i_a}{N_i}
\left(1-\frac{k^i_a N_i/t^i_a}{4\pi\ita_i}\right)
            + \order{\abs{q_i}}.  
\end{align} 
The minimization condition Eq.~\eqref{eq-dtauiU} 
has the trivial solution, i.e. $U_a^\pr(X_a)=0$, 
as the simultaneous $n$ linear equations,   
which can be achieved only if $n \ge A$. 
Whereas the non-trivial solution can exist if $\mathrm{rank}\,C < A$.

At the minimum $U_a^\pr = 0$, 
the second derivatives of the potential are given by 
\begin{align}
\begin{pmatrix}
 \partial_{\rta_i}\partial_{\rta_j} & \partial_{\rta_i}\partial_{\ita_j} \\
 \partial_{\ita_i}\partial_{\rta_j} & \partial_{\ita_i}\partial_{\ita_j}  
\end{pmatrix}
  U(X) &\ 
  \\ \notag 
  &\   \hspace{-3.0cm} = 
  \sum_a 
   \log \left(1+ \frac{\alpha_a}{X_a}\right) 
\begin{pmatrix}
 (\partial_{\rta_i} X_a)(\partial_{\rta_j}X_a) & 
 (\partial_{\rta_i}X_a)(\partial_{\ita_j}X_a) \\
 (\partial_{\ita_i}X_a)(\partial_{\rta_j}X_a) & 
 (\partial_{\ita_i}X_a)(\partial_{\ita_j}X_a)   
\end{pmatrix}. 
 \end{align}
The $2n\times 2n$ mass matrix for the modulus is rank $A$. 
Thus $n \ge A$ is necessary to 
make $\ita_i$'s massive at $\order{\abs{q}^0}$.   
Altogether, 
we find that the minimum $U^\pr_a=0$ can stabilize all of the modulus 
with $n$ massless modes only if $n=A$.
For $A > n$, 
the minimization condition 
may be satisfied in the non-trivial way, $U_a^\pr(X_a)\ne 0$ 
and the massless modes will obtain masses from the CW potential~\footnote{
The case of $(n,A) = (1,2)$ is discussed in Ref.~\cite{Higaki:2024jdk}, 
where one of the vector-like pair dominates the other 
and the leading $\rta$ dependence is generated from the lighter one 
with the suppression of $\order{q^{1+t^1_2/N}}$.  
}. 
In this work, we will focus on the case of $n=A$ as the minimal example 
to stabilize multiple moduli.

The simplest example is that $C_{ia}$ is a diagonal matrix, 
i.e. the normalized mass can be written as 
\begin{align}
 X_a(\tau) = X_a(\tau_a) = (2\ita_a)^{k_a} \abs{\Ykr{k_a}{1_{t_a}}(\tau_a)}^2.    
\end{align}
Hence, the normalized potential can be written as 
\begin{align}
 U(\tau) = \sum_a U_a(X_a(\tau_a)),  
\end{align}
and the mass of the modulus $\tau_a$ is generated only from $U_a(\tau_a)$.  
Thus in this case, the result in the single modulus case 
can be directly applied for each sector labeled by $a$.

Now, let us discuss the non-diagonal case. 
The trivial solution $U^\pr_a(X_a) = 0$ reads as 
\begin{align}
\label{eq-trivial22}
\frac{\mu^2}{\Lambda_Q^2} = X_a(\tau) + \order{\alpha_a}.   
\end{align}
This implies that 
the masses of the $A$ vector-like pairs are nearly degenerate at the minimum. 
To proceed, we assume the universal modular weight $k^i := k^i_a$ for $a$ for simplicity. 
In this case, 
\begin{align}
 X_a(\tau) =  \prod_i 
          (2\ita_i)^{k^i} \abs{\Ykr{k^i}{1_{t_a^i}}(\tau_i)}^2
       =  \prod_{i} \abs{c^{k^i}_{t^i_a}}^2 (2\ita_i)^{k^i} e^{-\frac{4\pi t^i_a}{N_i} \ita_i} 
           \left(1 + \order{q_i}\right)
\end{align}
Hence, the minimization condition requires that 
\begin{align}
\ell_a = \sum_i \left(\frac{4\pi t^i_a}{N_i}\right) \ita_i  =: \sum_i P_{ai}\; \ita_i,  
\quad 
\ell_a := 
 \log \left(
\frac{\Lambda_Q^2}{\mu^2} \prod_i (2\ita_i)^{k^i} \abs{c^{k^i}_{t^i_a}}^2 \right),  
\end{align}
for $a=1,2,\cdots,A$, so 
\begin{align}
\label{eq-Pinvl}
\sum_a \left[ P^{-1}\right]_{ia} \ell_a > 0 
\end{align}
is required for $\ita_i > 0$.  
Note that $\ell_a$ only logarithmically depends on $\ita_i$.  
For $n=2$, $N_1 = N_2 =:N$ and $c^{k^i}_{t^i_a} = 1$, 
this condition is simplified to 
\begin{align}
\label{eq-condSimplified}
 (t^1_1 - t^1_2)(t^2_2-t^2_1) > 0
\quad{\rm and}\quad 
 (t^1_1 - t^1_2) (t^1_1 t^2_2-t^1_2t^2_1) > 0.  
\end{align}
The solution is given by 
\begin{align}
 \ita_1 =&\ 
    -\frac{(k^1+k^2)N}{4\pi} \frac{t^2_1-t^2_2}{t^1_1t^2_2-t^1_2t^2_1}  
           \Wcal\left( 
    -\frac{4\pi}{(k^1+k^2)N} \frac{t^1_1t^2_2-t^1_2t^2_1}{t^2_1-t^2_2}  
           \left(\frac{t^2_2-t^2_1}{t^1_1-t^1_2}\right)^{\frac{k^2}{k^1+k^2}} 
           \left(\frac{\mu}{\Lambda_Q}\right)^{\frac{2}{k^1+k^2}} 
    \right), 
\notag \\ 
 \ita_2 =&\ 
    -\frac{(k^1+k^2)N}{4\pi} \frac{t^1_2-t^1_1}{t^1_1t^2_2-t^1_2t^2_1}  
           \Wcal\left( 
    -\frac{4\pi}{(k^1+k^2)N} \frac{t^1_1t^2_2-t^1_2t^2_1}{t^1_2-t^1_1}  
           \left(\frac{t^1_1-t^1_2}{t^2_2-t^2_1}\right)^{\frac{k^1}{k^1+k^2}} 
           \left(\frac{\mu}{\Lambda_Q}\right)^{\frac{2}{k^1+k^2}} 
    \right),  
\end{align}
and then 
\begin{align}
\label{eq-12ratio}
\frac{\ita_2}{\ita_1} = \frac{t^1_1-t^1_2}{t^2_2-t^2_1}.  
\end{align}
The qualitative features for more general cases 
will be similar to those obtained for this simple case.

\begin{figure}[t]
\begin{minipage}[t]{0.49\hsize} 
 \includegraphics[width=\hsize]{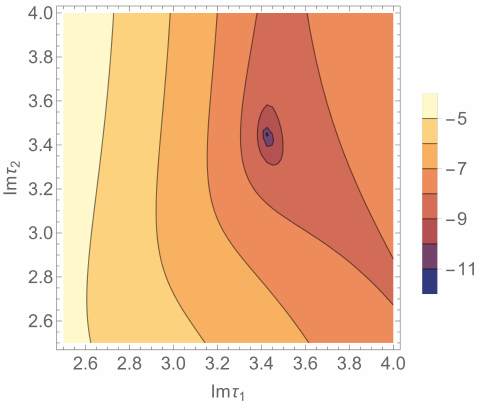} 
\end{minipage} 
\begin{minipage}[t]{0.49\hsize}
 \includegraphics[width=\hsize]{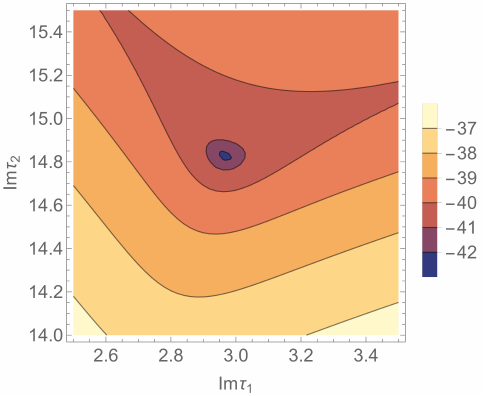} 
\end{minipage} 
\centering
\caption{\label{fig-2D}
$\log_{10}(U-U_{\mathrm{min}})$ on $(\ita_1, \ita_2)$ plane 
for $N=6$, $k=10$, $\alpha = 10^{-4}\times \mu^2/\Lambda_Q^2$.
The $\intZ^T_6$-charges are chosen as 
$(t^1_1, t^2_1, t^1_2, t^2_2) = (3,3,5,1)$ 
and $(0,3,5,2)$ on the left and right panels, respectively. 
$\mu/\Lambda= 10^{-1}$ ($10^{-9}$) in the left (right) panel.  
}
\end{figure}

Now we numerically show the existence of the minimum.  
We focus on the case of $N_1=N_2 = 6$, $k^i_a = 10$,  
where there are five singlet modular forms with $t = 0,1,2,3,5$. 
The normalization of the modular forms are chosen 
such that $c^{10}_{t} = 1$ with the universal scale factor $\Lambda_Q$. 
Figure~\ref{fig-2D} shows the potentials in the case 
of $(t^1_1, t^2_1, t^1_2, t^2_2) = (3,3,5,1)$ 
and $(0,3,5,2)$ on the left and right panels, respectively.  
In this figure, we choose $\alpha = 10^{-4}\times \mu^2/\Lambda_Q^2$ 
and $\mu/\Lambda_Q = 10^{-1}~(10^{-9})$ on the left (right) panel. 
On the left panel,  
the minimum is at $(\ita_1, \ita_2) \simeq  (3.43, 3.43)$, 
where $\ita_1/\ita_2 \simeq 1$ as expected from Eq.~\eqref{eq-12ratio}.  
The minimum is at  $(\ita_1, \ita_2) \simeq  (2.96, 14.8)$ 
where $\ita_1/\ita_2 \simeq 5$ on the right panel. 
Note that the largest ratio is $5 = N-1$ for $N=6$ in this setup, 
and the ratio can change from these values by changing  
the modular weights, normalizations of the modular forms and $N_i$ 
in more general cases.   
We have shown the simplest cases which can realize the two patterns 
of $\ita_i$. 
We set $\rta_i = 0$ for the numerical plots, 
but there is no visible difference for $\rta_i \ne 0$.

As for the single modulus case, 
the potential has the flat direction with $U^\pr_a=0$ which is approximately along the
$\rta_i$ directions for $\ita_i \gg 1$. 
For $n=A$ and the minimum $U^\pr_a = 0$ is realized, 
the normalized potential can be expanded around the minimum $\tau = \ita_0$ as 
\begin{align}
 U(X) \simeq  \frac{1}{2} \sum_a \log\left(1+\frac{\alpha_a}{X_a(\tau_0)}\right) 
               (X_a(\tau)-X_a(\tau_0))^2  
      \sim \frac{1}{2}\sum_a \alpha_a X_{a}(\tau_0) \left( 
        \frac{X_a(\tau)}{X_a(\tau_0)} -1 
           \right)^2,
\end{align}
where the constant term is omitted and $\alpha_a \ll 1$ is assumed 
in the second equality. Expanding the modular form as 
\begin{align}
 \Ykr{k^i_a}{1_{t^i_a}}(\tau_i) = q^{t^i_a/N} (1 + b^i_a q_i + \order{q_i^2}),   
\end{align}
the leading $\rta_i$ dependence appear in the potential as 
\begin{align}
 V_{\CW}(\rta_i) \simeq \sum_a \frac{m_a^2 m_{Q_a}^2(\tau)}{8\pi^2}
      \left(\sum_i b^i_a \abs{q_i} \cos\left(2\pi \rta_i\right) \right)^2.  
\end{align}
Thus the leading potential for $\rta_i$ is estimated as 
\begin{align}
V(\rta_i) \sim \max_{a,j} \left[\frac{m_a^2 m_{Q_a}^2(\tau)}{4(1+\delta_{ij})\pi^2} 
               b_a^i b_{a}^j \abs{q_i q_j} 
               \cos{(2\pi \rta_i)} \cos{(2\pi \rta_j)}   \right]. 
\label{eq:Vaxion}
\end{align}  
If $\rta_i$ is identified as the QCD axion, 
there exists the scalar potential
\begin{align}
V_{\rm QCD} \sim -\Lambda_{\rm QCD}^4 \cos\left( \theta_0 + 2\pi  \sum_{a,i}  \frac{ t^i_a }{N_i}\rta_i\right)
, 
\end{align}
where $\theta_0 ={\cal O}(1)$ and $\Lambda_\QCD$ is the QCD scale.
Hence, the massless mode is stabilized by the QCD effect and can solve the strong CP problem if 
\begin{align}
 \Delta \theta_\QCD \sim&\ \frac{m_a^2 m_{Q_a}^2}{4\pi^2 \Lambda_{\mathrm{QCD}}^4} 
              b^i_a b^j_a \abs{q_i q_j} 
\label{eq:Deltatheta}
\\ \notag
  \sim&\  10^{-10} \times b^i_a b^j_a  
                        \left(\frac{m_a}{10^{7}\;\GeV}\right)^2 
                        \left(\frac{m_{Q_a}}{10^{9}\;\GeV}\right)^2 
                        \left(\frac{0.2\;\GeV}{\Lambda_{\QCD}}\right)^4 
                        \left(\frac{\abs{q_i q_{{\mathrm{max}}}}}{10^{-45}}\right),  
\end{align}
where $q_{\mathrm{max}} = \mathrm{max}_{i} (q_i)$,  
is smaller than the experimental bound $\Delta \theta_{\rm QCD} < 10^{-10}$~\cite{Abel:2020pzs}. 
The critical value $\abs{q_i q_j} \sim 10^{-45}$  
corresponds to $\ita_i + \ita_j \simeq  16$. 
Note that $\rta_i$ with the smallest potential will be identified as the QCD axion.
Hence, the case of the right panel in Fig.~\ref{fig-2D} 
with $(\ita_1, \ita_2) \simeq (2.96, 14.8)$ 
could resolve the strong CP problem
and $\rta_2$ is predominantly the QCD axion. 
If $A>n$, then the minimum is not at the one with $U^\pr_a \ne 0$.  
The potential for $\rta_i$ will be suppressed only by $\order{q_i}$, 
and thus $\abs{q_i} \lesssim \order{10^{-45}}$, i.e. $\ita_i \gtrsim 16$ 
is necessary to be identified as the QCD axion.

\section{Conclusion}
\label{sec:conclusion}

In this paper, 
we study the multiple mouduli stabilization scenario 
by the radiative corrections from the masses invariant 
under direct products of the finite modular symmetries $\Gamma_{N_1}\times \cdots \Gamma_{N_n}$.
We showed that we need at least $n$ independent mass terms 
to stabilize $n$ moduli
associated with the direct product group.
There are $2n-A$ massless modes
at the minimum $U^\pr_a(X_a) =0$, 
where $A$ is the number of the mass terms generating the CW potential. 
These massless directions could obtain their masses dominantly 
by the non-perturbative effects in the same way as the QCD axion. 
The stabilization by the CW potential tends to have light direction 
due to the residual $\intZ^T_N$ symmetry which is a good symmetry 
for $\ita\gg 1$. 
In general, the potential along $\rta$ direction is suppressed 
by $\abs{q} = e^{-2\pi \ita}$ compared with that along $\ita$ direction. 
The flat direction could be identified as the QCD axion 
to solve the strong CP problem 
if $\abs{q}$ is sufficiently small, $\ita \gtrsim 15$.

In the minimal case $n=A$, 
there is only the trivial solution satisfying $U^\pr_a(X_a) =0 $ for $a=1,2,\cdots,n$, 
see Eq.~\eqref{eq-dtauiU}. 
In addition, the condition in Eq.~\eqref{eq-Pinvl} should be satisfied 
for $\ita_i > 0$. 
We explicitly studied the case of $n=2$ under the additional assumptions 
to simplify the condition Eq.~\eqref{eq-Pinvl} to Eq.~\eqref{eq-condSimplified} 
which constrains $\intZ^T_N$ charges. 
We showed explicit examples that have the minimum 
at both $\ita_1 \sim \ita_2$ and $\ita_1 \ll \ita_2$.  
The former case would be applicable to explain 
both hierarchical flavor structure and spontaneous CP violation~\cite{Abe:2023ilq,Kikuchi:2023fpl}. 
The latter could be applicable to address the flavor hierarchy 
and the strong CP problem in a model. 
It will be straightforward to extend the scenario to $n=3$, 
so that all of the three problems are addressed simultaneously. 
In such case, the strong QCD problem would be addressed by combination of the QCD axion 
and vanishing QCD $\theta$ angle before the symmetry breaking due to the spontanesous CP violation.
Note that $n=3$ may be motivated from toroidal compactification 
$T^2\times T^2 \times T^2$, 
where each torus provides a modulus.

In this work, we focused on the stabilization of the moduli 
which could be applied for models to address the problems in the SM. 
In addition, it will be interesting to study 
cosmological consequences of the moduli and axion fields, 
as studied for the simplest case in Ref.~\cite{Jung:2024bgi}, 
which are expected to be much lighter than the fundamental scale of a model. 
Regarding the multiple moduli setup, 
it will also be interesting to study models with the Siegel modular group 
based on $\mathrm{Sp}(2g,\intZ)$, $g\in\mathbb{N}$, instead of $\mathrm{SL}(2,\intZ)$.

\section*{Acknowledgement} 

This work was supported in part JSPS KAKENHI Grant Numbers No. JP22K03601 (T.H.),
JP23K03375 (T.K.) and JP24KJ0249 (K.N.).
The work of J.K. was supported by IBS under the project code, IBS-R018-D1.

\appendix

\section{Modular forms of $\Gamma_6$}
\label{appendix:gamma6}
We show $\Gamma_6$ modular forms~\cite{Li:2021buv}. First, we define Dedekind eta function $\eta(\tau)$ below:
 \begin{equation}
 \eta(\tau) := q^{1/24}\prod_{n=1}^{\infty}(1-q^n),\quad q:=e^{2\pi i\tau}.
 \end{equation}
 We define the following  functions from $\eta(\tau)$:
 \begin{eqnarray}
 Y_1(\tau)&=&3\frac{\eta^3(3\tau)}{\eta(\tau)}+\frac{\eta^3(\tau/3)}{\eta(\tau)},\nonumber
 \\
  Y_2(\tau)&=&3\sqrt{2}\frac{\eta^3(3\tau)}{\eta(\tau)},\nonumber
  \\
  Y_3(\tau)&=&3\sqrt{2}\frac{\eta^3(6\tau)}{\eta(2\tau)},\nonumber
  \\
  Y_4(\tau)&=&-3\frac{\eta^3(6\tau)}{\eta(2\tau)}-\frac{\eta^3(2\tau/3)}{\eta(2\tau)},\nonumber
  \\
  Y_5(\tau)&=&\sqrt{6}\frac{\eta^3(6\tau)}{\eta(2\tau)}-\sqrt{6}\frac{\eta^3(3\tau/2)}{\eta(\tau/2)},\nonumber
  \\
  Y_6(\tau)&=&-\sqrt{3}\frac{\eta^3(6\tau)}{\eta(2\tau)}+\frac{1}{\sqrt{3}}\frac{\eta^3(\tau/6)}{\eta(\tau/2)}-\frac{1}{\sqrt{3}}\frac{\eta^3(2\tau/3)}{\eta(2\tau)}+\sqrt{3}\frac{\eta^3(3\tau/2)}{\eta(\tau/2)}.
 \end{eqnarray}
 Then we can construct the singlet modular forms of $\Gamma_6$,
\begin{align}
    \Ykr{2}{1_1}(\tau) = -\frac{1}{\sqrt{6}} (Y_3 Y_6 - Y_4 Y_5), 
\end{align}
for $k=2$, 
\begin{align}
    \Ykr{4}{1_0}(\tau) = Y_1 (Y_1^3+2\sqrt{2}Y_2^3), 
    \quad
    \Ykr{4}{1_2}(\tau) = \Ykr{2}{1_1}(\tau)^2, 
\end{align}
 for $k=4$, 
 \begin{align}
    \Ykr{6}{1_0}(\tau) =&\ (Y_1Y_4-Y_2Y_3)\left\{
     3(Y_1Y_6-Y_2Y_5)^2 - (Y_1Y_4-Y_2Y_3)^2 
      \right\}, 
      \\ \notag 
    \Ykr{6}{1_1}(\tau) =&\ \Ykr{2}{1_1}(\tau) \Ykr{4}{1_0}(\tau), 
    \quad 
     \Ykr{6}{1_3}(\tau) = \Ykr{2}{1_1}(\tau) \Ykr{4}{1_2}(\tau), 
 \end{align}
 for $k=6$, and 
 \begin{align}
    \Ykr{10}{1_0}(\tau) =&\ \Ykr{4}{1_0}(\tau) \Ykr{6}{1_0}(\tau), 
    \quad 
     \Ykr{10}{1_1}(\tau) = \Ykr{4}{1_0}(\tau) \Ykr{6}{1_1}(\tau), 
      \quad 
     \Ykr{10}{1_2}(\tau) = \Ykr{4}{1_2}(\tau) \Ykr{6}{1_0}(\tau), 
      \\ \notag 
     \Ykr{10}{1_3}(\tau) =&\ \Ykr{4}{1_0}(\tau) \Ykr{6}{1_3}(\tau), 
      \quad 
     \Ykr{10}{1_5}(\tau) = \Ykr{4}{1_2}(\tau) \Ykr{6}{1_3}(\tau), 
 \end{align}
 for $k=10$. 
 These for the other weights are not shown here since we do not use them in this work. 
 Note that there are ambiguities of their normalizations. 
Here, we employ the normalization that the leading order in the $q$-expansion is simply given by 
\begin{align}
    \Ykr{k}{1_t}(\tau) \sim q^{t/6}\left(1 + \order{q} \right),  
\end{align}
i.e. $c^k_t = 1$ in Eq.~\eqref{ex-qexp}. 

\bibliography{refererences}
\bibliographystyle{JHEP}

 \end{document}